\documentstyle[prl,aps,twocolumn]{revtex}

\begin{document}\draft
\flushbottom
\twocolumn[
\hsize\textwidth\columnwidth\hsize\csname @twocolumnfalse\endcsname

\preprint{CU-TP-845; cond-mat/9707202}

\title{Novel Phenomena in Charged Bose Liquid}

\author{Kimyeong Lee and Oleg Tchernyshyov}
\address{Department of Physics, Columbia University, New York, NY
10027, USA}
\date{\today}
\maketitle 
\tightenlines
\widetext
\advance\leftskip by 57pt
\advance\rightskip by 57pt

\begin{abstract}
We investigate charged Bose liquid immersed in uniform background
charge at zero temperature.  Novel phenomena, such as oscillatory
shielding of external localized electric charge, rotons and charge
density waves (charge stripes in two dimensions), occur in any
dimensions.  Oscillatory shielding is caused by mixing between scalar
boson exchange and Coulomb interactions, which mediate opposite
forces. On the other hand, rotons and charge density waves are due to
attractive local self-interaction of bosons.  Rotons can be regarded
as a finite size charge density wave packet without any back flow. We
also comment on charge stripes observed recently in cuprates and
nickelates.
\end{abstract}


]

\narrowtext
\tightenlines

Recently there have been several experimental observations of charge
and spin stripes in high temperature superconductor
materials~\cite{st1,st2,st3}. There have been several theoretical
discussions about the origin and implications of these
stripes~\cite{zachar}. Experiments show on-set of charge order before
spin order, showing that charge stripe is primary and spin stripe is
secondary. The superconducting transition temperature is either zero
in compounds with a static wave of charge density, or significantly
reduced when dynamical correlations are observed.  It would be
interesting to find a simple model which produces charge stripes and
dynamical correlations.

In this paper we investigate charged Bose liquid immersed in uniform
external electric charge at zero temperature.  We describe charged
Bose liquid by a complex scalar field with nonlinear local
self-coupling and gauge coupling with photons. By analyzing linear
fluctuations around homogeneous ground state, we show that novel
collective modes, like oscillatory shielding of external localized
charges, rotons and charge density waves, occur in any dimensions.  We
identify physical reason behind these phenomena. Our model can be
regarded as a simple toy model which provides such rich phenomena, and
may also shed some light on current experiments.

Since unique is the rest frame of uniform external charge, there is no
Lorentz or Galilean symmetry in our theory. There have been many works
on this system with attention on vortex dynamics~\cite{klee1,strato}.
Vortex dynamics is particularly simplified in the self-dual
model\cite{klee1}, where magnetic flux vortices are shown not to carry
any intrinsic angular momentum.

The presence of uniform background charge leads to nontrivial
consequences. The average charge density of bosons is determined by
the condition of electric charge neutrality rather than by the
chemical and potential energy. For this reason, the (short-range)
self-interaction of bosons does not have to be repulsive for the
system to be stable. If this interaction is weakly attractive,
low-energy excitations from a uniform ground state are fluctuations of
charge density with a characteristic wave-length. These excitations
can be called rotons in analogy with those in superfluid helium. As
the strength of attraction increases, the frequency of this mode
vanishes and the uniform ground state becomes unstable, forming charge
density waves. In superfluid helium nonlocal attractive
self-interaction of helium atom plays a crucial role in the existence
of rotons. (For a review of current understanding of rotons in
superfluid helium, see Ref.~\cite{donnelly}.)

A closely related question is the screening of a probe charge. It is
normally expected that a localized external charge is fully shielded
by charged bosons; the amount of total charge inside a sphere with the
external charge in its center falls off exponentially with the
radius. However, in the case we study, the shielding can be
oscillatory. This originates from mixing of two opposite forces
mediated by scalar boson exchange and Coulomb interactions.

We find a link between oscillatory shielding and the existence of a
roton dip in the excitation spectrum. As stressed by
Feynman~\cite{feynman1}, the roton minimum is associated with a
short-range crystal-like structure in liquid helium. In that case, the
density-density correlation $\langle \rho({\bf r})\rho(0)\rangle$ is a
non-monotonic function of the distance, an effect similar to the
oscillatory screening of charge.

Within the scope of our model, a roton wave packet can be visualized
as a localized charge density wave. As discussed above, it originates
from the attractive self-interaction of bosons.  While the wave packet
itself is at rest (zero group velocity), charge density waves are
moving at the roton phase velocity.  Curiously, there is no net flow
of charge associated with this motion and therefore it is unnecessary
to attach a vortex ring to this wave packet in order to produce a back
flow. The idea of back flow goes back to Feynman and
Cohen~\cite{feynman2}, who tried to modify Feynman's original ansatz
to enforce the conservation of current. In our theory the charge
current is exactly conserved from the start and so there is no need
for back flow.  At least our result is consistent with the current view
of rotons in superfluid helium~\cite{donnelly}.

We describe charged Bose liquid by a bosonic scalar field $\psi$ with
a local self-interaction and the electromagnetic coupling. This liquid
is immersed in a uniform background charge $\rho_e>0$.  The Lagrangian
for charged Bose liquid is
\begin{eqnarray}
{\cal L} = & & -\frac{1}{16\pi}F_{\mu\nu}F^{\mu\nu} +
i\psi^*(\partial_0-iqA_0)\psi \\ 
& & - \frac{1}{2m}|(\partial_i - iqA_i )\psi|^2 - U(|\psi|^2) - 
q\rho_e A_0,  
\end{eqnarray}
where $\hbar=c=1$.  The Gauss law constraint is
\begin{equation}
\partial_i F_{0i} + 4\pi q(\rho  -\rho_e) = 0.
\end{equation}
As the charged background is static, homogeneous and uniform,
there are conserved energy  and  linear and angular momenta. The Noether
angular momentum density is not gauge invariant and needs
a modification, which plays a crucial role in vortex
dynamics~\cite{klee1}.

The homogeneous configuration satisfying the field equations is
given as 
\begin{equation}
 \langle A_\mu \rangle=0\,\,\,\, {\rm and}\,\,\,\,
\langle \psi \rangle = \sqrt{\rho_e} e^{-i\alpha t},
\label{homo}
\end{equation}
where $\alpha= dU/d\rho |_{\rho_e}$.  Total charge density is equal to
zero and the system is charge neutral. We approach this problem by the
semiclassical approximation around this uniform background.  To find
out elementary excitations on this homogeneous configuration, we study
linear fluctuation equations in the unitary gauge ${\rm Arg}\,\psi =
0$, and so $\langle A_0 \rangle= \alpha/q$. In the Fourier mode
$\delta \psi, \delta A_\mu \sim e^{-i\omega t +ip^ix^i} $, the
linearized field equations become
\begin{eqnarray}
& & -(\frac{p^2}{2m}+2\lambda\rho_e)\delta\psi + q\sqrt{\rho_e}\delta
A_0=0,
 \nonumber\\ 
& & p^2\delta A_0 + \omega p^i\delta A_i +
8\pi q\sqrt{\rho_e}\delta \psi=0, \nonumber\\ 
& & (\omega^2-p^2-m_A^2)\delta A_i + p^i p^j \delta A_j + \omega  p^i
\delta A_0=0,
\end{eqnarray}
where $\lambda=d^2 U/d\rho^2|_{\rho_e}$ and $m_A^2=4\pi\rho_eq^2/m$.
The parameter $\lambda$ is the strength of the local self-coupling for
linear fluctuations.

From these equations, we get the dispersion relation for the
transverse modes,
\begin{equation}
(\omega^2-p^2-m_A^2)\epsilon_{ijk}p^jA^k=0.
\end{equation}
This Anderson-Higgs mechanism explains the Meissner effect with the
London penetration length $\lambda_L = 1/m_A$.  The dispersion
relation for the mixed mode of $\delta \psi,\delta A_0$ and $ p^i
\delta A_i$ is
\begin{equation}
\omega^2= m_A^2+\frac{p^2}{2m} \left(\frac{p^2}{2m}+2\lambda\rho_e \right).
\label{disper}
\end{equation}
For the time-independent fluctuations, we get the static relation
\begin{equation}
p^2= -2m\lambda \rho_e \pm 2m \sqrt{(\lambda\rho_e)^2-m_A^2}.
\label{disper1}
\end{equation}
This describes how the system responds to any time-independent
external source.  

Depending on the range where the parameters of the theory lies, we
will see that several collective modes are allowed by
Eqs.~(\ref{disper}) and (\ref{disper1}), which are summerized as
follows:
\begin{eqnarray}
\lambda\rho_e>m_A;     & &\,\, {\rm exponential}\,\,{\rm  shielding},
\label{exp} \\ 
|\lambda  \rho_e|<m_A; & &\,\, {\rm oscillatory}\,\,{\rm shielding},
\label{oscil} \\ 
0>\lambda \rho_e >-m_A; & &\,\, {\rm rotons},
\label{roton}\\ 
\lambda \rho_e< -m_A;   & &\,\, {\rm charge}\,\,{\rm  density}\,\,{\rm  waves},
\label{cdw} \\
\lambda \rho_e<< -m_A;  & &\,\, {\rm Wigner}\,\,{\rm  crystal}. 
\label{wigner}
\end{eqnarray}

When the electromagnetic coupling constant $e^2$ vanishes, this theory
has a smooth limit. It describes a neutral Bose liquid with local
self-interaction.  We know well that with repulsive self-interaction
$\lambda>0$, the homogeneous configuration is stable with positive
pressure. Elementary excitations in that case are massless phonons
with sound speed $v^2_s = \lambda \rho_e/m$.  If we have introduced a
proper nonlocal self-interaction as in superfluid helium, the roton
mode would have appeared. There is no massive mode but there exists a
natural length scale, i.e., the coherence length $\xi =
1/\sqrt{4m\lambda\rho_e}$.  With attractive self-interaction
$\lambda<0$, the configuration is unstable. Bosons aggregate and form
droplets, or Q-balls, with mean charge density $\rho_c$, at which
$U(\rho)$ takes the minimum value.

Let us first study how the system responds to the static localized
probe charge. Far away from this source, the field configuration
deviates little bit from the uniform configuration~(\ref{homo}). Since
the charge couples to $A_0$, Eq.~(\ref{disper1}) would govern the
behavior of such a deviation.  There are three interesting ranges of
the parameters from Eq.~(\ref{disper1}).  When the coupling is
strongly repulsive so that the parameters lies in the
range~(\ref{exp}), the solution of Eq.~(\ref{disper1}) is pure
imaginary. Thus the local source is shielded exponentially. However,
the solution of Eq.~(\ref{disper1}) becomes complex when the parameter
lies in the range~(\ref{oscil}).  Thus the fields would fall off
exponentially but the sign of the field would oscillate, like the
amplitude of underdamped harmonic oscillators. We will call this as
{\it oscillatory shielding}.

To find the physical origin of such oscillatory screening, let us
recall the old lore.  For two like  sources in large separation,
the virtual scalar boson exchange leads to the attractive Yukawa force, but
the virtual photon exchange leads to the repulsive Coulomb force.  For
the opposite charges, the scalar force is repulsive and the Coulomb
force is attractive. This can be seen directly by obtaining the
interacting potentials from the corresponding Lagrangians,

\begin{eqnarray}
& & {\cal L}_1 = -\frac{1}{2}(\partial_i f)^2 - \frac{1}{2} m_1^2 f^2
+ \rho(x) f,
 \\ & & {\cal L}_2 = \frac{1}{2}(\partial_i A_0)^2 +
\frac{1}{2}m_2^2 A_0^2 + \rho(x) A_0,
\end{eqnarray}
where $\rho(x) = q_1 \delta^3(x-x_1)+ q_2\delta(x-x_2)$. Here we have
neglected the time dependence and scaled the field variables. The sign
of the field propagator is also associated with the sign of the
exchange force. Thus the $A_i$ exchange would lead to the similar
force as the scalar boson exchange. We already know this is true as
two parallel currents are attractive and two opposite currents are
repulsive. (The magnetic flux of vortices is not a direct source of
$A_i$ unlike the current.  The force between fluxes by the $A_i$
exchange is more like the Coulomb case.)  When there is a mixing
between $A_0$ and $f$ by a Lagrangian ${\cal L}_3 = \mu^2 f A_0$, the
corresponding static relation of the total Lagrangian ${\cal
L}_1+{\cal L}_2 +{\cal L}_3$ becomes
\begin{equation}
(p^2+m_1^2)(p^2+m_2^2)+\mu^4=0.
\label{m12}
\end{equation}
When $2\mu^2>|m_1^2-m_2^2|$ so that the mixing is strong enough, the
solution $p^2$ changes from negative real numbers to complex
number. This is how oscillatory shielding appears.

More interestingly, when the parameters lie in the range~(\ref{cdw}),
the solution of Eq.~(\ref{disper1}) becomes real numbers. The
fluctuations would not die out. Especially near $\lambda
\rho_e \approx -m_A$, the charge density would be oscillating with a
characteristic wave length
\begin{equation}
l_{CDW}= \frac{2\pi}{\sqrt{2mm_A}}.
\label{lcdw}
\end{equation}
This describes {\it charge density waves}. (In two dimensions, it
would be {\it charge stripes}.) In our case $m_1^2 $ and $m_2^2$ of
Eq.~(\ref{m12}) can be negative and so the charge density wave mode
is possible.

Let us now consider the time-dependent fluctuations described by
Eq.~(\ref{disper}). The slope of $\omega(p)$ at $p=0$ becomes negative
when the self-interaction is attractive so that the parameters lie in
the range~(\ref{roton}).  It has the minimum at
\begin{equation}
p_* = \sqrt{2m(-\lambda\rho_e)},
\end{equation}
with the value
\begin{equation}
\omega_* = \sqrt{m_A^2 - \lambda^2\rho_e^2}.
\end{equation}
As long as $\lambda \rho_e>-m_A$, the mode at the minimum has the
positive energy and is called {\it rotons}.

For a roton mode, we have to choose a particular direction, ${\bf
p}_*$. We can form the roton wave packet by superposing the plane waves
with an envelop function ${\cal F}$.  Since the dispersion relation
distinguishes the parallel and orthogonal momenta with respect to ${\bf
p}_*$, ${\cal F}$ can depend on parallel and orthogonal momenta
independently.  Especially if ${\cal F}$ is peaked at ${\bf p}_*$, the
group velocity would be zero. 

Now one may wonder what happens to charge density and current in the
roton wave packet, especially whether there exists any {\it back
flow}. We first note that our theory works in any spatial
dimension. Thus we first consider the system in one dimensional
space. When the wave packet size is large, the major contribution to
charge density and current comes from the ${\bf p}_*$ plane
wave. Inside the wave packet, both positive and negative charge
regions of wave length $\lambda_*= 2\pi/p_*$ moves forward with phase
velocity $v_*=\omega_*/p_*$. Thus, there is no net charge transfer
from one end to another end of the wave packet, and no need for back
flow of charge. This is consistent with the fact that the current
conservation holds for linear fluctuations.  The charge density wave
inside the roton wave packet is exactly like sound wave in fluid,
which does not transport any mass density and needs no back flow.

Coming back to the two or three dimensional problem, we note that the
envelop function can be chosen to be Gaussian along the parallel and
orthogonal directions, independently. Then the above one-dimensional
argument shows that there is no back flow even in these cases.  If
there is any back flow, it may manifest itself as vortex ring, or
smoke ring around the roton wave packet, which is clearly
nonperturbative and contradicts the perturbative nature of our
fluctuations. Smoke ring is also impossible in one-dimensional space.

If the interaction between bosons become more attractive so that
Eq.~(\ref{cdw}) holds, the roton mode becomes unstable and charge
density waves appear. The homogeneous field configuration would be no
longer the ground state configuration. Since the charge density wave
mode is characterized by a single length scale $l_{CDW}$, the ground
state is made of a {\it static} plane wave of the given wave length,
pointing a particular direction, say $\hat{\bf z}$. This lowers the
gradient energy. Thus the translation symmetry is broken along the z
axis but is invariant along the x-y plane. The rotational symmetry is
also partially broken to the z directional rotation. There would be a
Goldstone boson mode which modulate the position of the crest of
charge density waves. The amplitude of charge density waves would be
fixed by the nonlinear interaction. The exact semiclassical
description of the ground state needs the evaluation of the full
nonlinear field equations, which will tell us whether the background
charge is exposed at the bottom of the wave. (This may tell us whether
this phase is superconducting or metallic along the z-direction).

When the attractive interaction becomes still stronger, the unstable
modes can have a range of momentum. We  have to treat the full
nonlinear field equation to find the exact ground state field
configuration.  However, one can guess the ground state configuration
in this case by starting from the other end where the electromagnetic
coupling is negligible. If we turn off the electromagnetic coupling,
charged bosons will aggregate to many droplets, or Q-balls, and then
they will grow. When we turn on the electromagnetic coupling, they
cannot grow forever. Rather Q-balls of suitable size would arrange
themselves into a crystal structure to lower the Coulomb energy. This
configuration is analogous to the {\it Wigner crystal} of electrons in a
uniform positive charged background. In this case the translation and
rotational symmetries are broken to a discrete crystal group.

The recently discovered {\it charge stripes} and {\it dynamical
correlations} can be regarded as charge density waves and rotons in
our context.  While their dynamical origin may be far complicated,
charged Bose liquid considered here may capture some physics (aside
from spin and lattice effects). As the dopping increases in these
materials, the carrier charge density changes.  There are
several possibilities of behavior. Bosons can be repulsive or
attractive in small density, though they are expected to be repulsive
in high density. 

At very low density $m_A\sim \sqrt{\rho_e}\gg| \lambda\rho_e|$ and
therefore the charge shielding is oscillatory with the wave vector
$p\approx (i\pm 1)\sqrt{mm_A}$.  Experimentally it may be difficult to
tell this type of shielding from purely exponential because the
imaginary part of $p$ equals the real part.  If $\lambda$ is positive
and approximately constant, the exponential screening appears in high
density.

Second possibility is that $\lambda$ is negative in small density and
so the roton mode presents in low density.  As the density $\rho_e$
increases, the roton dip becomes more prominent.  Being lowest energy
excitations of the system, rotons deplete the condensate density at
temperatures equal to or above their energy, like  dynamical
correlations in real material. In higher density the repulsive
interaction turns on and the roton mode disappears.

If the attractive interaction remains to high enough densities, the
roton mode becomes unstable at some critical density and the system
may not be superfluid any longer.  At a higher density, the reduced
attraction can no longer maintain a static charge density wave and
superfluidity comes back. At the moment of instability the wavelength
of charge density waves is determined by the boson mass $m$ and the
photon mass $m_A$ (the London penetration depth).  If we take $m$
equal to two electron masses and the London length $\lambda_L = 2000
\AA$, a typical experimental value, the resulting CDW wavelength
(\ref{lcdw}) is of order
\begin{equation}
l_{CDW} = \pi\sqrt{\frac{\lambda_L\hbar}{m_e c}} = 8.5 \rm \AA.
\end{equation} 
This is of the same order as the experimental value $\sim 15 \AA$,
which is encouraging.  (Note that this analysis is bases on linear
fluctuations and so the Meissner effect is still true. But this may
not be true once the exact semiclassical ground state is found.)

It is surprising that our simple system shows many fascinating
phenomena, as oscillatory shielding, rotons and charge density waves.
Our work suggests that these phenomena are quite generic.  We now have
tools to explore quantum mechanics of these collective modes.  We can,
for example, calculate spin of rotons in our model.  It would be also
interesting to analyze the system by using the method of many body
physics.  It is known that when $\lambda=0$ the homogeneous
configuration changes to the Wigner crystal in high enough density due
to the Coulomb repulsion~\cite{ceperl}, which we do not see in our
approach.

Even though  we have not considered here, there exist also magnetic flux
vortices in the theory.  The naive division between type I and type II
should be redrawn by studying the long-distance interaction between
magnetic flux vortices  in the exponential shielding region.

It would be most interesting to find out whether our consideration has
any direct bearing on real superconductors, whose physics was
investigated in context of charged Bose liquid~\cite{mott}.  Recently
there has been some proposal of more realistic field theoretic models
of the BCS superconductors at zero temperature\cite{stone}.  It would
be interesting to find out whether our consideration applies to that
case. There are also numerous works relating superconductivity and
charge density wave in one-dimensional systems\cite{gruner}.  We hope
that our work can provide some useful insights
here as well.

\noindent {\bf Acknowledgment:} The work of K.L. was supported in part
by the NSF Presidential Young Investigator Fellowship. The work of
O.T. was supported in part by the US Department of Energy and Japan
NEDO. The part of this work was done in Aspen Center for Physics.  We
thank Elihu Abrahams, Sidney Coleman, David Ceperley, Miklos Gyulassy,
T.D. Lee, John Negele, Michael Stone, and Tomo Uemura for useful
discussions.


\begin{references}

\bibitem{st1}{C.H. Chen, S.-W. Cheong and A.S. Cooper,
Phys. Rev. Lett. {\bf 71}, 2461 (1993); J.M. Tranquada et al.,
Phys. Rev. B {\bf 52}, 3581 (1995); V. Sachan et al., Phys. Rev. B
{\bf 51}, 12742 (1995).}

\bibitem{st2}{J.M. Tranquada et al., Nature {\bf 375}, 561 (1995);
Phys. Rev. B {\bf 54}, 7489 (1996); J.M. Tranquada et al., Phys. Rev. Lett.
{\bf 78}, 338 (1997).}

\bibitem{st3}{S-W Cheong et al., Phys. Rev. Lett. {\bf 67}, 1791
(1991); T.E. Mason, G. Aeppl and H.A. Mook, Phys. Rev. Lett. {\bf 68},
1414 (1992); T.R. Thruston et al., Phys. Rev. B {\bf 46}, 9128 (1992).}


\bibitem{zachar}{For some of recent theoretical discussions, see O. Zachar,
S.A. Kivelson and V.J. Emery, {\it Landau Theory of Stripe Phases in
Cuprates and Nickelates}, cond-mat/9702055; C. Nayak and F. Wilczek,
Int. J. Mod. Phys. B{\bf 10}, 2125 (1996).}




\bibitem{klee1}{K. Lee, Phys. Rev. D{\bf 49}, 4265  (1994).}

\bibitem{strato}{G.N. Stratopoulos and T.N. Tomaras, Physica D{\bf
89}, 136 (1995); M. Stone, {\it Magnus and other forces on vortices in
superfluids and superconductors}, to appear (1997).}


\bibitem{donnelly}{R.J. Donnelly, Phys. World
{\bf 10} {\rm no.} 2, 25 (1997).}

\bibitem{feynman1}{R.F. Feynman, Phys. Rev. {\bf 91}, 1301 (1953);
Phys. Rev. {\bf 94}, 262 (1994).}

\bibitem{feynman2}{R.F. Feynman and M. Cohen, Phys. Rev. {\bf 102},
1189 (1956).}





\bibitem{ceperl}{D.M. Ceperley and B.J. Alder, Phys. Rev. Lett. {\bf 45},
566 (1980).}

\bibitem{mott}{A.S. Alexandrov and N.F. Mott,  Rep. Prog. 
Phys. {\bf 57}, 1197 (1994).} 


\bibitem{stone}{M. Stone, Int. Jour. Mod. Phys. B{\bf 9}, 1359 (1995);
I. R. Aitchison, P. Ao, D. J. Thouless and X.-M. Zhu, Phys. Rev. B{\bf
51}, 6531 (1995).}


\bibitem{gruner}{G. Gruner, {\it Density waves in solids},
(North-Holland, New York, 1994); S. Kagoshima, H. Nagasawa,
T. Sambongi, {\it One-dimensional conductors} (Springer-Verlag,
Berlin, 1988).}


\end{references}
\end{document}